\documentclass[twocolumn,showpacs,amsmath,amssymb,prl]{revtex4}

\begin{document}

\title{The Work-Hamiltonian Connection and the Usefulness of the
  Jarzynski Equality for Free Energy Calculations \footnote{This
    article has been submitted to the Journal of Chemical Physics,
    which can found at \url{http://jcp.aip.org}. }}
\author{Eric N. Zimanyi} \author{Robert J. Silbey}
\affiliation{Department of Chemistry, Massachusetts Institute of
  Technology, Cambridge, Massachusetts 02139}

\begin{abstract}
  The connection between work and changes in the Hamiltonian for a
  system with a time-dependent Hamiltonian has recently been called
  into question, casting doubt on the usefulness of the Jarzynski
  equality for calculating free energy changes. In this paper, we
  discuss the relationship between two possible definitions of free
  energy and show how some recent disagreements regarding the
  applicability of the Jarzynski equality are the result of different
  authors using different definitions of free energy.  Finally, in
  light of the recently raised doubts, we explicitly demonstrate that
  it is indeed possible to obtain physically preprintrelevant free
  energy profiles from molecular pulling experiments by using the
  Jarzynski equality and the results of Hummer and Szabo.

\end{abstract}


\maketitle

Single molecule experiments, such as the stretching of a polymer
molecule using an atomic force microscope or laser tweezers, have
become common in the last decade \cite{bustamante,harris}. The goal is
often the determination of the free energy surface along some
coordinate of the molecular potential energy surface. In order to
proceed, one invokes the Jarzynski equality using the extension
derived by Hummer and Szabo \cite{hum_szab}.

Although there has been some controversy about these theoretical
advances, it is fair to say that their use in interpreting nano-scale
single molecule experiments is widespread. Thus any question that they
may be fundamentally in error must be carefully examined.

Recently, questions have been raised about the connection between work
and changes in the Hamiltonian for a system with a time-dependent
Hamiltonian, casting doubt on the applicability of the Jarzynski
equality for computing free energy changes \cite{failure}. Here, we
discuss these questions and show that the Jarzynski equality can be
usefully applied to determine physically relevant free energy changes.

Consider a system with Hamiltonian $\mathcal{H}_0(x)$, where $x$
represents the microstate of the system, and suppose that this system
is subject to a time-dependent force $f(t)$ acting along some
coordinate $z(x)$. From the perspective of classical mechanics, we
have two options for treating the force. We may consider it as an
external force not included in the Hamiltonian of the system and study
the evolution of a system governed by $\mathcal{H}_0(x)$ under the
effect of the external force $f(t)$ acting along $z(x)$.
Alternatively, we may include the force in the Hamiltonian of the
system and study the evolution of a system governed by
$\mathcal{H}(x,t) = \mathcal{H}_0(x) -z(x) f(t)$.

In the first case, we are considering a time-independent Hamiltonian
under the effect of an external force $f(t)$. According to classical
mechanics, the work done by the external force up to time $\tau$ is
\begin{equation}
  W(\tau) = \int_{t=0}^{t=\tau} f(t) d\{z[x(t)]\}
\end{equation}
and we have the usual result that the work done on the system equals
its energy change,
\begin{equation}
  \mathcal{H}_0[x(t_2)]-\mathcal{H}_0[x(t_1)] = W(t_2) - W(t_1).
\end{equation}
The free energy change appropriate for this first description of the
system is
\begin{equation}
  G(z_2) - G(z_1) = - \log\left[ \frac{\int d x \delta[z(x) - z_2]e^{-\mathcal{H}_0(x)}}{\int d x\delta[z(x)-z_1] e^{- \mathcal{H}_0(x)}}\right]
\end{equation}
($k_B T=1$ throughout).

In the second case, we consider the time-dependent Hamiltonian
$\mathcal{H}(x,t) = \mathcal{H}_0(x) - z(x) f(t)$. In this description
of the system, $f(t)$ is an internal force and there should be no
expectation that the work done by $f(t)$ equals the change in energy
of the system.  Here we consider the \textit{thermodynamic work},
\begin{equation}
  W_t(\tau) = \int_0^\tau dt \frac{\partial \mathcal{H}}{\partial t},
\end{equation}
which by definition equals the change in energy of the system.  The
appropriate free energy change to consider for this description of the
system is
\begin{equation}
  \label{eq:j_freechg}
  \Delta G_t(\tau) = G_t(\tau) - G_t(0) = - \log\left[ \frac{\int d x e^{-\mathcal{H}(x,\tau)}}{\int d x e^{- \mathcal{H}(x,0)}}\right].
\end{equation}
We note, as suggested by Vilar and Rubi (VR) \cite{failure}, that this
second description of the system is not unique -- adding a term $g(t)$
to the Hamiltonian has no effect on the dynamics of the system but
changes the values of $W_t$ and $\Delta G_t$.

In considering the effect of a force $f$ on a harmonic spring of force
constant $k$, VR describe the system according to the first picture
and obtain $\Delta G = W = f^2/2k$ while Horowitz and Jarzynski (HJ)
use the second picture and obtain $\Delta G_t = W_t = - f^2/2k$
\cite{j_comment,reply_j}. Both of these results are correct in their
respective descriptions, and mean different things. In particular, VR
are describing the free energy change associated with changing the
length of the spring in the absence of an external force; the force is
only a tool used to measure the free energy profile of the free
spring. Meanwhile, HJ are describing the free energy change of the
combined force-spring system as a function of the force.

The Jarzynski equality is framed in the second of our descriptions and
expresses a relation between $W_t$ and $\Delta G_t$ \cite{jarz},
\begin{equation}
  e^{-\Delta G_t} = \langle e^{-W_t}\rangle.
\end{equation}
The validity of this expression is not in question -- only its utility
in describing free energy changes in a system. VR point out that
$\Delta G_t$ depends on the arbitrary choice of $g(t)$ in the
Hamiltonian and leads to arbitrary free energy changes. If all that
can be extracted from the Jarzynski equality is this arbitrary $\Delta
G_t$, then the Jarzynski equality seems to be of little use. We shall
show, however, that this is not the case.

Consider a single-molecule pulling experiment, for which the Jarzynski
equality has frequently been applied \cite{bustamante,harris}. In
studying the unfolding of a biomolecule, one is often interested in
the free energy profile $G(z)$ as a function of end-to-end distance
$z$. We could map the free energy by reversibly pulling the ends of
the molecule and measuring the work exerted by the external force as a
function of $z$. This is of course the classic method and corresponds
to VR's analysis of the harmonic spring.

We could also try to get the free energy profile using the Jarzynski
equality. Direct application of the Jarzynski equality to yield
$\Delta G_t$ gives the free energy difference between the free
molecule and the molecule with a certain force applied to it. This is
not in itself a particularly useful quantity and is not the free
energy profile. Hummer and Szabo have, however, shown how to obtain
free energy profiles from single-molecule pulling experiments
\cite{hum_szab}.

Consider an unperturbed system described by a Hamiltonian
$\mathcal{H}_0(x)$. When a time-dependent perturbation is applied
along some coordinate $z(x)$, we write the new Hamiltonian as
$\mathcal{H}_0(x) + \mathcal{H}'(z,t)$. Hummer and Szabo have shown
that the unperturbed free energy profile along coordinate $z$ can then
be reconstructed as
\begin{equation}
  \label{eq:humszab_orig}
  G(z_0) = - \log \langle \delta [ z(t) - z_0] e^{-W_t + \mathcal{H}' (z,t)}\rangle,
\end{equation}
where the average is over all trajectories of the system in the
presence of the perturbation \cite{hum_szab}.

We now apply this result to a macroscopic, deterministic spring and
show how the Jarzynski equality can be used to calculate $G(z)$,
thereby reconciling the results of HJ and those of VR. In this case,
there is only one degree of freedom so the microstate $x$ is simply
the length of the spring. Our model is $\mathcal{H}_0(x) = p^2 + k x^2
/2$ and $\mathcal{H}(x,t) = p^2 + k x^2 / 2 - f(t) x$, where $f(t)$
switches from 0 to $f_0$ over time $0<t<\tau$.  The pulling process is
finished at $t=\tau$, at which time $f(\tau)=f_0$ and from classical
mechanics $x(\tau) = f_0/k$.

Inserting this into Hummer and Szabo's result,
\begin{equation}
  \label{eq:humszab}
  G(z) = - \log \langle \delta [ x(\tau) - z] e^{-W_t - f_0 x(\tau)}\rangle,
\end{equation}
where we have used the fact that in this case our coordinate of
interest $z(x)$ is just $x$.

In this deterministic case only $z=f_0/k$ contributes to the average
and we have
\begin{equation}
  \label{eq:subz}
  G(z = f_0/k) = - \log \langle  e^{-W_t - f_0^2/k}\rangle\\ =- \log \left[ \langle  e^{-W_t}\rangle e^{-f_0^2/k} \right].
\end{equation}
For reversible pulling, HJ calculate that the work distribution is
sharply peaked at $W_t = -f_0^2/2k$, so we finally obtain
\begin{equation}
  G(z = f_0/k) = - \log \left[ e^{f_0^2/2k} e^{-f_0^2/k}\right] = + f_0^2/2k,
\end{equation}
which agrees with VR and is the expected result for the free energy
profile of a Hookean spring.

Consider the effect of adding an arbitrary $g(t)$ to the general
Hamiltonian. The effect on $\Delta G_t$ is easily seen from
Eq.~\eqref{eq:j_freechg} to be
\begin{equation}
  \Delta G_t^{new}(\tau) = \Delta G_t(\tau) + [g(\tau)-g(0)].
\end{equation}
Since the term $g(t)$ redefines the zero of energy at each point in
time, it is expected that $\Delta G_t(\tau)$ will be affected as it is
comparing free energies at two different times. Before ascribing a
physical interpretation to $\Delta G_t$, it must be corrected by
subtracting this arbitrary change in the zero of energy.

We now examine the effect of an arbitrary $g(t)$ on the free energy
profile $G(z)$ computed via Jarzynski's equality. We then have
\begin{eqnarray}
  \mathcal{H'}^{new}(x,t) &=& \mathcal{H'} (x,t) + g(t)\\
  W_t^{new} &=& W_t + g(\tau) - g(0)
\end{eqnarray}
and Eq.~\eqref{eq:humszab_orig} becomes
\begin{equation}
  G^{new}(z_0) = - \log \langle \delta [ z(t) - z_0] e^{-W_t + \mathcal{H}' (z,t)+g(0)}\rangle,
\end{equation}
which can be simplified to
\begin{equation}
  G^{new}(z_0) = G(z_0) - g(0).
\end{equation}
So adding a time-dependent term $g(t)$ shifts the overall free energy
profile $G(z)$ by an additive constant, but has no effect on relative
free energies.

We can consider the same situation from the perspective of
thermodynamics in one dimension. The internal energy of the system is
given by $U = \mathcal{H}_0$ and its enthalpy by the Legendre
transform $\mathcal{H}_0 - f z$. We can then define two free energies,
$G = U - ST$ and $G_t = U - fz - ST$.

The method of VR is constructed to measure $G$ as a function of
position, $G(z)$, while HJ are calculating $G_t$ as a function of $f$,
$G_t(f)$.  As long as the fluctuations in $x$ are small at a given $f$
(ie, we are in the thermodynamic limit), we can use the simple
relation $G(z) = G_t(f) + fz$ to convert between the two quantities.
Outside of this limit, there is not a simple relation between the two
quantities but the method of Hummer and Szabo discussed above can be
used to reconstruct $G(z)$ from pulling experiments.

In conclusion, we have shown that by properly applying the Jarzynski
equality, the textbook result for the free energy profile of a spring
is correctly recovered. More importantly in light of recent doubts, we
have reaffirmed the applicability of the Jarzynski equality to the
analysis of single-molecule pulling data.

\begin{acknowledgments}
  Part of this research was supported by the NSF under grant
  CHE0556268. One of the authors (E.Z.) acknowledges financial support
  from FQRNT.  We acknowledge an email exchange with Prof. Jarzynski.
\end{acknowledgments}

\end{document}